\def\bx{{\bf x}}
\def\br{{\bf r}}
\def\bR{{\bf R}}
\def\bk{{\bf k}}
\def\bK{{\bf K}}
\def\bG{{\bf G}}

\documentstyle[aps,eqsecnum,twocolumn]{revtex} 

\begin{document}

\draft

\title{The {\it GW} space-time method for the self-energy of large systems}
\author{Martin M. Rieger$^\S$\cite{auth0}, L. Steinbeck$^\dagger$
\footnote{Department of Physics, University of York, York YO10 5DD, U.K., 
          email: ls9@york.ac.uk, 
          Tel. (+44) (0) 1904 432208  Fax (+44) (0) 1904 432214},
        I. D. White$^\S$, H. N. Rojas$^\S$\cite{auth2}, 
        and R. W. Godby$^\dagger$\cite{auth3}}
\address{$^\S$Cavendish Laboratory, University of Cambridge, Madingley Road,
             Cambridge, CB3 0HE, UK}
\address{$^\dagger$Department of Physics, University of York, Heslington,
            York YO1 5DD, UK}

\date{\today}

\maketitle

\begin{abstract}
We present a detailed account of the $GW$ space-time method. The method
increases the size of systems whose electronic structure can be
studied with a computational implementation of Hedin's $GW$ approximation. At
the heart of the method is a representation of the Green's function $G$
and the screened Coulomb interaction $W$ in the real-space and imaginary-time 
domain, which allows a more efficient computation of the self-energy
approximation $\Sigma = iGW$. For intermediate steps we freely change
between representations in real and reciprocal space on the one hand, 
and imaginary time and imaginary energy on the other, using fast Fourier
transforms. The power of the method is demonstrated using the example
of Si with artificially increased unit cell sizes. \\
keywords: electronic structure, quasiparticle energies, self-energy 
calculations, GW approximation
\end{abstract}

% 71.15.Th: Electronic structure calculations - other methods
% 71.20.-b: Electron density of states and bandstructure of crystalline solids
% 79.60.Jv: Interfaces, heterostructures, nanostructures

\section{Introduction} 

Computational electronic structure theory for real materials depends on
the use of simplifying approximations for the many-electron problem. Two
successful approaches have been the use of density functional theory
and many-body perturbation theory. The density functional approach is
overwhelmingly dominated by the local-density approximation (LDA) of Kohn
and Sham \cite{KOS65}, and extensions thereof, such as gradient corrections
or self-interaction corrections. The many-electron problem is mapped onto an
effective non-interacting electron problem and solved for the ground-state
density and energy. The limitation of this approach lies in the fact that in
principle it gives no access to the excitation spectrum of the system under
study, even if approximations for the exchange and correlation potential
are further refined. This limitation is felt particularly severely in
semiconductor physics, where many of the phenomena of interest are centered
on properties of the excited states. For this class of materials Hybertsen
and Louie\cite{HYL85,HYL286} showed that the $GW$ approximation, first 
proposed by Hedin\cite{HED65} in 1965, allows computation of band gaps 
in remarkably good agreement with experiment for a series of semiconducting 
and insulating materials.

The $GW$ approximation gives a comparatively simple expression for the
self-energy operator, which allows the one-particle Green's function of an
interacting many-electron system to be described in terms of the Green's
function of a hypothetical non-interacting system with an effective potential.
The Green's function contains information not only about the ground state
density and energy but also about the quasiparticle spectrum. The $GW$
approximation has still proved computationally very expensive and has mainly
been used to determine the quasiparticle spectrum of bulk semiconductors and
insulators \cite{HYL286,GSS86,ZHL91}, although progress has also been made in
the treatment of systems such as surfaces\cite{LOU94}, clusters\cite{ORG95} and
simple polymers\cite{EFZ96}. The issue of self-consistency has only begun to be
addressed fairly recently\cite{GBH95,BAH96,SHI96,SEG98}. 
The situation is similar for total energy
calculations, where thorough investigations so far have been made only for the
homogeneous electron gas\cite{HOB98}. First $GW$ calculations of the charge 
density of Si and Ge have been performed recently\cite{RIG98}.

The method we describe in this paper substantially reduces the computational
effort needed to study larger systems. The core idea was outlined in a Letter
by Rojas, Godby, and Needs\cite{RGN95}. Here we give a full account of the
method and detailed aspects of our implementation. The improvements in
efficiency over traditional implementations of the $GW$ approximation in a
reciprocal-space formalism result from choosing the representation most
suitable to the computational step being undertaken, either reciprocal space or
real space on the one hand and imaginary time or imaginary energy on the other,
and switching between representations easily with the help of fast Fourier
transforms (FFTs). The choice of representing the time/energy dependence on
the imaginary instead of on the real axis allows us to deal with smooth,
decaying quantities which give faster convergence. To obtain the self-energy
eventually on the real energy axis, we fit a model function to the computed
self-energy on the imaginary axis, and continue it analytically to the real
axis.

The paper is organised as follows: The second Section reviews briefly the
equations needed for implementation of the $GW$ approximation. Section
III describes the essential concepts of the $GW$ space-time method. In the
fourth Section we outline detailed aspects of the implementation such as
mesh discretisation and exploitation of symmetry. In Section V we discuss 
the analytic continuation procedure for the self-energy, and subsequent 
determination of the quasiparticle energies. In the sixth Section we 
present results for bulk silicon and silicon with artificially increased 
unit cell sizes, and discuss scaling behaviour for these examples. A summary 
concludes the paper. Atomic units are used throughout.

\section{The $GW$ approximation}

The central idea of Hedin's $GW$ approximation is to approximate the
self-energy operator $\Sigma$ by
\begin{equation}
\label{SIGMA}
\Sigma({\bf r},{\bf r}';\omega) = \frac{i}{2\pi} \int\limits_{-\infty}^{\infty} 
                          d\omega'\; W({\bf r},{\bf r}';\omega')\;
                          G({\bf r},{\bf r}';\omega+\omega') e^{i\omega'\delta},
\end{equation}
where $\delta$ is an infinitesimally small positive time, $W$ is the
screened Coulomb interaction,
\begin{equation}
\label{W}
W({\bf r},{\bf r}';\omega) = \int d^3{\bf r}'' v ({\bf r} - {\bf r}'')\;
                         \epsilon^{-1}({\bf r}'',{\bf r}';\omega) 
\end{equation}
where $v ({\bf r}-{\bf r}'')$ is the Coulomb interaction $1/|{\bf r}-{\bf r}''|$
and $G$ is the one-particle Green's function. $G$ itself depends on $\Sigma$
through the Dyson equation and should arguably be determined self-consistently.
In practice, however, in almost all calculations for real systems $G$ has been
approximated by the non-interacting Green's function at the LDA level, i.~e.,
\begin{equation}
\label{GLDA}
G^{LDA}({\bf r},{\bf r}';\omega) = \sum\limits_{n\bk} 
         \frac{\Psi_{n\bk}(\br)\; \Psi_{n\bk}^*(\br')}
              {\omega-\epsilon_{n\bk}-i\eta}
\end{equation}
where $\eta$ is a positive (negative) infinitesimal for occupied (unoccupied)
one-particle states. The wavefunctions $\Psi_{n\bk}$ in this equation are 
eigenfunctions, with eigenvalues $\epsilon_{n\bk}$, determined from a 
self-consistent LDA calculation for the system under consideration.

For the inverse dielectric function in Eq.\ (\ref{W}) one has to rely on
suitable approximations. We use the random phase approximation (RPA),
\begin{eqnarray} \label{EPSRPA}
 \epsilon^{RPA}(\br,\br';\omega) &=& \delta(\br-\br') \\ \nonumber
 &-& \int d\br''v(\br-\br'')\; P^0(\br'',\br';\omega)
\end{eqnarray}
with the irreducible polarization propagator $P^0$ at RPA level given by
\begin{eqnarray} \label{P0RPA}
\lefteqn{ P^0(\br,\br';\omega)}   \\  \nonumber 
  && = -\frac{i}{2\pi} \int\limits_{-\infty}^{\infty} 
       d\omega' \;G^{LDA}(\br,\br';\omega') \;G^{LDA}(\br,\br'; \omega'-\omega).
\end{eqnarray}
Part of the efficiency of our method derives from the fact that the
convolutions Eqs.\ (\ref{SIGMA}) and (\ref{P0RPA}) in the frequency domain
become simple multiplications in the time domain (real or imaginary): see 
next section. For real times the Green's function Eq.\ (\ref{GLDA}) becomes
\begin{eqnarray}
\lefteqn{
G^{LDA}(\br,\br';\tau) } && \\[12pt]
 && = \left\{\begin{array}{ll}
 \phantom{-}
 i \sum\limits_{n\bk}^{occ}\Psi_{n\bk}(\br)\Psi^*_{n\bk}(\br')
   \exp(-i\epsilon_{n\bk}\tau) , & \ \tau < 0, \\[12pt]
-i \sum\limits_{n\bk}^{unocc}\Psi_{n\bk}(\br)\Psi^*_{n\bk}(\br')
   \exp(-i\epsilon_{n\bk} \tau) , & \ \tau > 0.
\end{array}
\right. \nonumber
\end{eqnarray}
For imaginary times our expression for $G^{LDA}$ (see Eq.\ (\ref{GLDAST})
later) corresponds to analytically continuing the $\tau < 0$ form (the
retarded Green's function) to the positive imaginary time axis, and the $\tau >
0$ form (the advanced Green's function) to the negative imaginary axis.

Once the self-energy operator is known we can employ first-order perturbation
theory in $\left\langle \Sigma -V_{\text{xc}}^{\text{LDA}}\right\rangle$
to compute quasiparticle corrections to the LDA eigenenergies (see Section
\ref{QPECORR}).

\section{The $GW$ space-time method}

\label{GWST}

\subsection{Mathematical formulation}

The traditional way to set up and solve the equations for the $GW$
approximation has been to express and compute all quantities in the
reciprocal-space and energy domain, using a plasmon-pole model for the energy
dependence of $W$ \cite{HYL286}. To compute $P^0$ (Eq.\ (\ref{P0RPA}))
and $\Sigma$ (Eq.\ (\ref{SIGMA})) this involves sums scaling with the
fourth power of the number of plane waves $N_\bG$ used to represent the
wavefunctions and quadratically with the number of energy points $N_\omega$
used to represent the energy dependence, whereas the scaling in the real
space and time domain is quadratic in $N_\bG$ and linear in $N_\omega$,
since convolutions in the reciprocal-space and energy domain become simple
multiplications in the real-space and time domain. The expressions for
$\epsilon$ and $W$, on the other hand, are more efficiently calculated in a
reciprocal-space and energy representation. As fast Fourier transforms permit
us to switch between representations with a numerical cost that scales like
$N\log{N}$, where $N$ is the number of points involved in the FFT, we can
efficiently exploit the advantages offered by the one or other representation.

The time- or energy-dependence of the quantities involved shows a structure
that is not easily represented on an equidistant grid suitable for an
FFT. However, we can rigorously analytically continue the quantities to
the imaginary time or energy axes where the structure is much smoother
and therefore amenable to a representation on an equidistant grid. This
point is illustrated by Fig.~1 of Ref.\ \onlinecite{RGN95} which shows the
energy dependence on the real and imaginary axis of the imaginary part of
the self-energy $\Sigma(k,\omega)$ of jellium with a density parameter of 
$r_s=2.0$. The rather ragged shape on the real energy axis contrasts
with the smooth shape on the imaginary energy axis. We emphasise that, while
the same amount of physical information is contained for $\Sigma$ or similar
functions on the real or imaginary time or energy axis, to obtain a well 
converged final answer for quantities that go through several Fourier 
transformations, the important information is more easily represented 
in imaginary time/energy, in much the same way that the choice of basis set 
can reduce the effort needed to satisfactorily represent wavefunctions. 
This is also shown in plots of the self-energy in Section \ref{RESULTS}, 
where it can be seen that the smooth form of the self-energy on the 
imaginary axis still allows stable and accurate reproduction of the more 
complicated behaviour on the real axis.

Using the imaginary time/energy representation allows us to explicitly
take the time- or energy-dependence into account without having to rely on
a plasmon-pole or other model for most of the calculation.  Only after
the full imaginary-energy dependence of the expectation values of the
self-energy operator has been established do we use a fitted model function
(whose sophistication may be increased as necessary with negligible expense),
which we then analytically continue to the real energy axis to compute the
quasiparticle energies.

The Fourier transforms between the complex axes work like their counterparts
on the real axes, except that additional factors of $\pm i$ have to be
included:
\begin{equation}
F(i\tau )=\frac i{2\pi }\int\limits_{-\infty }^\infty d\omega F(i\omega )
          \exp (i\omega \tau ),
\end{equation}
\begin{equation}
F(i\omega )=-i\int\limits_{-\infty }^\infty d\tau F(i\tau )\exp (-i\omega \tau).
\end{equation}
Mathematically they can be understood as Laplace transforms followed by
analytic continuation to the imaginary axis.

The computational steps which are successively undertaken in the $GW$
space-time method are in detail:
\begin{enumerate}

\item
Construction of the Green's function in real space and imaginary
time:\cite{Foot1}
\begin{eqnarray}
\label{GLDAST}
\lefteqn{
G^{LDA}(\br,\br';i\tau) } && \\[12pt]
 && = \left\{\begin{array}{ll}
 \phantom{-}
 i \sum\limits_{n\bk}^{occ}\Psi_{n\bk}(\br)\Psi^*_{n\bk}(\br')
   \exp(\epsilon_{n\bk}\tau) , & \tau > 0, \\[12pt]
-i \sum\limits_{n\bk}^{unocc}\Psi_{n\bk}(\br)\Psi^*_{n\bk}(\br')
   \exp(\epsilon_{n\bk}\tau) , & \tau < 0, \nonumber \\ \\
\end{array}
\right. 
\end{eqnarray}

\item
formation of the RPA irreducible polarizability in real space and imaginary
time:
\begin{equation}
\label{P0}
P^0(\br,\br';i\tau) = -iG^{LDA}(\br,\br';i\tau)G^{LDA}(\br',\br;-i\tau), 
\end{equation}

\item
Fourier transformation of $P^0$ to reciprocal space and imaginary energy and
construction of the symmetrised dielectric matrix\cite{BAR86} in reciprocal
space,
\begin{eqnarray}
\label{symdielm}
\lefteqn{
\tilde\epsilon(\bk,\bG,\bG';i\omega) = \delta_{\bG\bG'} } && \nonumber \\
&& {}-\frac{4\pi}{\left|\bk+\bG\right|\left|\bk+\bG'\right|}
      P^0(\bk,\bG,\bG';i\omega),
\end{eqnarray}

\item
inversion of the symmetrised dielectric matrix for each $\bk$ point and each
imaginary energy in reciprocal space,

\item
calculation of the screened Coulomb interaction in reciprocal space:
\begin{eqnarray}
W(\bk,\bG,\bG';i\omega) & = &
\frac{4\pi}{\left|\bk+\bG\right|\left|\bk+\bG'\right|} \nonumber \\[12pt]
& \times & \tilde\epsilon^{-1}(\bk,\bG,\bG';i\omega),
\end{eqnarray}

\item
Fourier transformation of $W$ to real space and imaginary time,

\item
computation of the self-energy operator:
\begin{equation}
\Sigma(\br,\br';i\tau) = iG(\br,\br';i\tau)W(\br,\br';i\tau),
\end{equation}

\item
evaluation of the expectation values:
\begin{equation}
\left<\Psi_{n\bk}|\Sigma(i\tau)|\Psi_{n\bk}\right>,
\end{equation}

\item
Fourier transformation of the expectation values to imaginary energy,

\item
fitting of a model function to the expectation values of the self-energy,
allowing analytic continuation onto the real energy axis,

\item
evaluation of the quasiparticle corrections to the LDA eigenvalues
by first-order perturbation theory in 
$\left\langle \Sigma -V_{\text{xc}}^{\text{LDA}}\right\rangle $. 
\end{enumerate}

\subsection{Discretisation of the equations}
\label{DISCRETIS}

A practical implementation on a computer requires the integrals described in
the last section to be discretised and suitably truncated. Exploitation of
symmetry can help to keep the computational cost for real materials down. 
These issues will be addressed in this section.

The quantities we are dealing with in real space, such as $G$, $P$, $W$,
and $\Sigma$ are non-local operators that decay as their two spatial
arguments move apart. With the exception of $W$, which will receive some
extra attention, the non-locality is short-ranged, i.e, decaying faster
than $|\br-\br'|^{-2}$.  This suggests that the distance the two variables
are allowed to move apart can be restricted suitably and the functions can
be assumed zero beyond this range.  We call this range the interaction cell
(IC). Furthermore, by the symmetry of the crystal, one of the arguments
of any of these operators, which shall be symbolically denoted here by
$F(\br,\br')$, can be restricted to the irreducible wedge of one unit cell
(IUC). The coarseness of the grid and the size of the interaction cell
determine the precision. A shape of interaction cell which is compatible
with the fast Fourier transforms and preserves symmetry at the same time is
the Wigner-Seitz cell of a lattice whose defining vectors are multiples of
the primitive vectors of the crystal lattice. We call this lattice the IC
lattice (ICL)  (see Figure \ref{grids}). 
We choose equal spacing for the IUC and IC grids. They
could be offset from each other which would avoid the singularity of the
Coulomb potential in real space but requires the handling of additional phase
factors. We therefore normally choose the two grids to have no offset between 
them and deal with a single real-space grid (RSG), which is defined by vectors
which are integer fractions of the primitive lattice vectors.
The treatment of the singularity of the Coulomb potential is discussed
in sections \ref{DIELM} and \ref{SCRCOUL} below.

To summarise: the crystal is defined by the three primitive vectors
${\bf a}_1$, ${\bf a}_2$, ${\bf a}_3$, the ICL by the three vectors
\begin{equation}
\label{BFL}
{\bf l}_i = N^{\bR}_i {\bf a}_i,\hskip 12pt i = 1,2,3,
\end{equation}
and the RSG by three vectors
\begin{equation}
\label{BFS}
{\bf s}_i = {\bf a}_i/N^{\br}_i,\hskip 12pt i = 1,2,3.
\end{equation}
In the example shown in Fig.\ \ref{grids} the $N^{\bR}_i$ are 2 and the 
$N^{\br}_i$ are 3.

The grid vectors $\bx$ in the IC are integer linear combinations of the 
${\bf s}_i$ and fulfil the Wigner-Seitz condition
\begin{equation}
\label{IC}
\bx\cdot{\bf L} \le \frac12 L^2,
\end{equation}
for any vector ${\bf L}$ that is an integer linear combination of the vectors
${\bf l}$. Only one of possibly several vectors for which the equality in Eq.\
(\ref{IC}) holds and which differ only by a lattice vector of the ICL must be
contained in the IC, or, expressed differently, only half the surface of the
Wigner-Seitz cell defined by Eq.\ (\ref{IC}) is part of the IC.
The integers $N^\bR_i$ in Eq.\ (\ref{BFL}) determine the shape and 
size of the interaction cell in real space and the k-point grid in reciprocal 
space, as explained below. If there is no reason
to believe that the non-locality of the operators has a very different range
in one direction than in another, they should be chosen in such proportion
to each other that the shape of the IC is as close to a sphere as possible.

The first argument $\br$ of the functions of type $F$ (see above) is again an
integer linear combination of the vectors ${\bf s}$ and can be restricted to
the irreducible wedge of one unit cell of the crystal lattice. This unit cell
can in principle have any shape, but it helps to avoid having to deal with
phase factors when applying symmetry operations if this unit cell is also
chosen to have Wigner-Seitz shape, i.~e.,
\begin{equation}
\br\cdot\bR \le \frac12 R^2,
\end{equation}
where $\bR$ is any crystal lattice vector and of any two $\br$ which are
connected by a symmetry operation in the crystal's space group only one is
kept.

The choice of lattices in real space uniquely determines conjugate lattices
in reciprocal space and vice versa, if one follows the rule that the discrete
approximation to the Fourier transform must leave the functions unchanged
if applied forwards and backwards in succession. In reciprocal space,
functions of type $F$ (see above) are functions of the two variables $\bG$
and $\bK$, where $\bG$ is a reciprocal lattice vector and $\bK$ can be
written as $\bK = \bk + \bG'$, where $\bG'$ is again a reciprocal lattice
vector and $\bk$ a reciprocal vector in the first Brillouin Zone (BZ). $\bK$
is restricted to the Wigner-Seitz cell of the reciprocal lattice of the RSG,
which we shall call simply the {\em reciprocal-space cell} (RC) and $\bG$
is restricted to its irreducible wedge (IRC). The spacing of the $\bk$
is determined by the primitive vectors of the reciprocal ICL and we shall
call the grid that is defined that way the $\bK$ grid (KG)\cite{FootR1}.

The transformation from real-space to reciprocal-space representation of $F$ is
given by
\begin{eqnarray}
\lefteqn{F(\bG,\bK) =} \\ \nonumber
&& \frac1{N_{IC}}\sum_{\br}^{IUC}\sum_{\rho}^{star(\br)}\sum_{\br'}^{IC(\br)}
F(\br,\br') e^{-i\rho\br\cdot\bG} e^{i\rho(\br'-\br)\cdot\bK}.
\end{eqnarray}
$\rho$ denotes a point group element (including possibly a non-primitive
translation in non-symmorphic symmetry groups) and the sum over $\rho$ runs
over all symmetry operations that generate the full star of $\br$, i.~e., the 
set of all RSG points in the Wigner-Seitz cell that are obtained by applying 
symmetry operations of the crystal to this particular RSG point $\br$ in the 
IUC. $N_{IC}$ is the number of RSG points contained in the IC. 
The reverse transformation is given by
\begin{eqnarray}
\lefteqn{F(\br,\br') =} \\ \nonumber
&& \frac1{N_{UC}}\sum_{\bG}^{IRC}\sum_{\rho}^{star(\bG)}\sum_{\bK}^{RC}
F(\bG,\bK) e^{i\rho\br\cdot\bG} e^{-i\rho(\br'-\br)\cdot\bK}.
\end{eqnarray}
The sum over $\rho$ runs here over all symmetry operations that generate the
full star of $\bG$ and $N_{UC}$ is the number of RSG points contained in one
unit cell.

Because of the symmetry of the problem and the fact that $\bK = \bk +\bG'$,
$F$ can in reciprocal space be alternatively written as a square matrix
$F_{\bk}(\bG,\bG')$ with $\bk$ restricted to the irreducible wedge of
the Brillouin Zone and $\bG$ and $\bG'$ given everywhere in the RC. $\bk$
is here written as an index to emphasise the matrix nature of $F$ in this
representation.

It should be briefly mentioned that the `natural' cell form for a
three-dimensional FFT is a parallelepiped and not a Wigner-Seitz cell such as
we are dealing with. However, there is a unique mapping between the two shapes
using translations by vectors of the lattice defining the Wigner-Seitz cell in
question, because $F$ is invariant under such translations by construction.

Apart from the real-space and reciprocal-space representations we will
occasionally use a mixed-space representation. This is defined by
\begin{equation}
\label{FK}
F_\bk(\br,\br') = \sum_\bR^{IC} F_\bR(\br,\br')\exp{(-i\bk\cdot\bR)},
\end{equation}
with the reverse transformation
\begin{equation}
F_\bR(\br,\br') = \frac{1}{N_\bR}\sum_\bk F_\bk(\br,\br')\exp{(i\bk\cdot\bR)}.
\end{equation}
When we use the mixed-space representation, $\br$ and $\br'$ are always
understood to be confined to the UC (or its irreducible wedge, see below). The
notation $F_\bR(\br,\br')$ is another way of writing \mbox{$F(\br+\bR,\br')$}
and has been chosen to emphasise that any point on the RSG in the IC can be
written as the sum of a crystal lattice vector $\bR$ and a vector in the UC.
The sum in Eq.\ (\ref{FK}) runs over all crystal lattice vectors in the IC. The
$\bk$ are vectors in the first BZ of the crystal and are the conjugate vectors
of the $\bR$. Symmetry can be exploited to reduce the number of points needed
to represent $F$ in mixed space. For example, $\br$ can be restricted to the
irreducible wedge of the UC while $\br'$ is given for every point of the RSG in
the UC and $\bk$ on every point of the KG within the first BZ, or $\bk$ can be
restricted to the irreducible wedge of the first BZ, with $\br$ then given on
every RSG point in the UC.

A mixed-space (MS) formalism was used by Blase et al. \cite{BRL95} for
calculating the polarizability and dielectric function of periodic systems. A
mixed-space representation was also used and described by Godby, Sham and
Schl\"uter \cite{GSS88} for $GW$ calculations, albeit without referring
to it by that name. Blase et al.\cite{BRL95} discuss the computational
efficiency of the MS scheme in comparison with a direct real-space (RS)
approach. In Fig.~2 of Ref.\ \onlinecite{BRL95} they show how the number
of $({\bf r},{\bf r}')$ pairs which have to be computed in order to set
up the polarizability $P^0_{\bf k} ({\bf r},{\bf r}')$ for one ${\bf k}$
in the MS method (using a $4\times4\times4$ Monkhorst-Pack ${\bf k}$
grid) depends on the range of interaction (non-locality range) $R_{max}$.
They compare this to the number of $({\bf r},{\bf r}')$ pairs obtained
by confining $|{\bf r}-{\bf r}'|$ to a sphere of radius $R_{max}$ which
is assumed to be the corresponding number of pairs to be computed in a RS
method. This is somewhat misleading since -- as is discussed above -- the
size of the IC in real space is determined by the size of the ${\bf k}$
grid of the conjugate mixed- or reciprocal-space quantity. Hence it does
not make sense to increase the interaction sphere beyond the boundaries
of this IC which correspond to a non-locality range of $R_{max}$=14.5 a.u.
for a $4\times4\times4$ ${\bf k}$ grid. Furthermore, the real-space quantity
$P^0 ({\bf r},{\bf r}')$ contains information on all ${\bf k}$ points. So
for a proper comparison between the two methods the number of pairs given
in Ref.\ \onlinecite{BRL95} for the MS approach has to be multiplied by
the number of special ${\bf k}$ points which is 10 for a $4\times4\times4$
Monkhorst-Pack grid. The statement made in Ref.\ \onlinecite{BRL95} that
the calculation of $P^0 ({\bf r},{\bf r}',\omega)$ requires a double BZ
summation in the real-space scheme - as opposed to a single BZ summation
in the MS scheme - does not hold any more if the Green's function is set
up in mixed space and then Fourier transformed to real space, as described 
in section \ref{GREENPOL} below.

So far we have established the nature of the grids in real and reciprocal
space that are suitable for use with the discrete Fourier transform. These
grids fill a Wigner-Seitz-cell shaped volume which in turn can be
uniquely mapped onto a parallelepiped, the shape that is required by the
discrete Fourier transforms.  However, the starting point of the $GW$
calculation is the output of a standard LDA calculation using plane waves
and pseudopotentials which determines the Fourier coefficients of the
wavefunctions $\Psi_{n\bk}(\bG)$ for all reciprocal lattice vectors that
lie within a sphere defined by $(\bk+\bG)^2$ less than some cutoff. The RC
must therefore be big enough to comprise all of these `shifted spheres'. The
volume of the smallest possible such cell will be typically
2 to 4 times larger than the volume of the individual spheres. Furthermore,
to strictly prevent aliasing in the steps where we replace a convolution in
one representation by a multiplication in the conjugate representation,
we would have to choose the grid for the FFTs twice as large as this
minimal cell in every dimension, increasing the volume by a factor of eight.
This means that the FFT grid would have to comprise between 16 and 32 times
more points than the initial LDA calculation had reciprocal lattice vectors.
We have found that in practice aliasing effects are very small once we
choose enough plane waves for good overall convergence, and in the limit of
an infinite number of plane waves any aliasing effects vanish strictly along
with any other error caused by truncation of the set of reciprocal lattice
vectors. Therefore it would not be justified to accept the huge overhead
imposed by the doubling of the grid size in all dimensions, and we usually
restrict our FFT grid to the minimum Wigner-Seitz cell as defined above,
leaving us with an FFT grid with between 2 and 4 times as many points as
the LDA calculation used reciprocal lattice vectors.

The computational effort can be further reduced by physical considerations. The
FFT grid fills a Wigner-Seitz cell shaped volume in either space. However, if
we assume in real space that the range of non-locality is uniform in all
directions we can neglect all the points in the interaction cell outside the
largest possible sphere inscribed into it. Similarly we can usually assume that
the Fourier coefficients in reciprocal space fall to zero after an equal length
in all directions and thus we can cut back in reciprocal space to the largest
possible sphere inscribed into the RC lattice. It has to be kept in mind that
for symmetry reasons it is the vectors $\bk+\bG$ which must fit into the sphere
and not merely the vectors $\bG$.

\section{Numerical aspects and scaling with system size}

\label{ASCALING}

As the $GW$ space-time method is primarily designed to enable larger
systems to be studied within the framework of the $GW$ approximation, in
this section we study the scaling of the computational cost with system
size. To gauge the cost of a calculation, we first look at the real-space,
imaginary-time representation. What matters here is the number of points
in the irreducible wedge of the unit cell, the number of points in the
interaction cell and the number of points on the imaginary time axis. We
will discuss the time dependence of the operators in the next section and
concentrate here on the spatial dimensions. Since in our setup the grid
spacing in the unit cell and in the interaction cell are chosen equal,
the three factors determining the cost of the calculation are the size of
the IUC, which is a system property, and the size of the interaction cell
and the grid spacing, which are convergence parameters.

\subsection{Green's function and polarizability}
\label{GREENPOL}

The setting up of the Green's function is the starting point for the $GW$
calculation. We take the Fourier coefficients of the lattice-periodic part of
the Bloch functions, $u_{n\bk}(\bG)$, and the eigenvalues $\epsilon_{n\bk}$
from a previous standard LDA calculation. The eigenvalues are expressed on an
energy scale with zero at the Fermi energy, which for semiconductors and
insulators is taken to be in the middle of the band gap. The wavefunctions
$u_{n\bk}$ are transformed via an FFT to real space and the Green's function
initially computed in mixed space:
\begin{eqnarray}
\label{GMIXED}
\lefteqn{
G^{LDA}_{\bk}(\br,\br';i\tau) } && \\[12pt]
&& = \left\{\begin{array}{ll}
\phantom{-}i\sum\limits_n^{occ}\Psi_{n\bk}(\br)\Psi_{n\bk}^*(\br')
            \exp(\epsilon_{n\bk}\tau),
& \tau > 0, \\[12pt]
-i\sum\limits_n^{unocc}\Psi_{n\bk}(\br)\Psi_{n\bk}^*(\br')
             \exp(\epsilon_{n\bk}\tau),
& \tau < 0. \nonumber \\ \\
\end{array}
\right.
\end{eqnarray}

The number of unoccupied bands included in the Green's function for $\tau < 0$
is a convergence parameter. Because of the rapid decay of the exponentials
for higher energies it suffices to set a cutoff energy for bands included
in the sum that is considerably smaller than the largest LDA eigenvalue,
but this cutoff must be several times larger than the highest quasi-particle
energy to be computed. We denote the total number of bands, occupied and
unoccupied, included in the sum as $N_{bands}$.

The Green's function can then be transformed from mixed space to real space
using an FFT:
\begin{equation}
G^{LDA}_\bR (\br,\br';i\tau) = \frac{1}{N_\bR}\sum_{\bk} G^{LDA}_{\bk}
                               (\br,\br';i\tau) \exp(i\bk\cdot\bR),
\end{equation}
where the sum goes over the {\bf k} grid in the Brillouin zone corresponding 
to the interaction cell (see III(B) earlier).

To set up the function in mixed space (Eq.\ (\ref{GMIXED})) we need
$N_{IUC}\times N_{UC}\times N_{bands}$ multiplications in the spatial
dimensions. The transformation to real space brings in another factor of
$N_{\bR}$, the number of unit cells contained in the interaction cell,
(disregarding the additional factor of $\log{N_{\bR}}$ from the FFT), so that
the overall scaling is like $N_{IUC}\times N_{IC}\times N_{bands}$. Since the
number of bands is determined by a cutoff energy that is constant for a given
material, $N_{bands}$ will grow linearly with system size. The computational
cost at this stage therefore scales quadratically with system size assuming a
fixed size of interaction cell (i.~e., fixed range of non-locality). Additional
care is necessary once the unit cell outgrows the range of non-locality. In
this case there is only a single $\bk$ point necessary and the transformation
between mixed and real space is redundant. To prevent a crossover to cubic
scaling the interaction cell must be kept constant at its converged size
and will then not be a multiple of the unit cell size, but smaller than it.

The next stage, where the irreducible polarizability is formed in real space
according to Eq.\ (\ref{P0}), scales linearly.

\subsection{Dielectric matrix}
\label{DIELM}

The formula for the dielectric function in real space and on the imaginary
energy axis reads
\begin{equation}
\label{EPSREAL}
\epsilon(\br,\br';i\omega) = \delta(\br-\br')
                 - \sum_{\br''}^{IC(\br')}v(\br-\br'')P^0(\br'',\br';i\omega).
\end{equation}
Here $\br'$ is restricted to the IUC, $\br$ to within an IC around $\br'$, and
$\br''$ in the sum on the right hand side runs over an interaction cell around
$\br'$. The right hand side contains a convolution that can be more efficiently
handled in reciprocal space where it becomes a simple multiplication. As only
one of the two spatial arguments of the dielectric function is involved, we
can write
\begin{equation}
\epsilon(\bK,\br') = v(\bK)P^0(\bK,\br'),
\end{equation}
where $\bK$ is the conjugate variable of $\bx = \br' - \br$ in reciprocal
space. (The imaginary-energy argument is suppressed here and until the end of 
this subsection.) Again we can see that for a given interaction
cell size the computational effort to set up the dielectric function scales
linearly with the number of points in the IUC.

The long-range behaviour of the dielectric function requires some special
consideration. Eq.\ (\ref{EPSREAL}) contains the convolution of the
long-ranged Coulomb potential with the short-ranged non-locality of $P^0$. In
fact, for any $\br'$
\begin{equation}
\label{INTP0}
\int d\br P^0(\br,\br') = 0,
\end{equation}
whereas
\begin{equation}
\label{INTV}
\int d\br v(\br) = \infty.
\end{equation}
The analytic convolution of the two functions over all space yields a function
whose integral over all space is finite, meaning that the Fourier coefficient
for $\bK = 0$ has a finite, non-zero value. To preserve this behaviour, which is
important if one is actually interested in computing the value of the
dielectric constant but also for fast convergence with respect to interaction
cell size of the quasiparticle energies, one has to use a modified Coulomb
interaction in Eq.\ (\ref{EPSREAL}). By construction, the property expressed by
Eq.\ (\ref{INTP0}) is within our discrete and strictly finite range
approximation compressed into the interaction cell:
\begin{equation}
\sum_{\br}^{IC(\br')} P^0(\br,\br') = 0.
\end{equation}
To obtain the right dielectric constant, we have to compress the property
expressed by Eq.\ (\ref{INTV}) into the finite interaction cell as well. A
natural way to do this is to use the reciprocal-space definition of $v$:
\begin{equation}
\label{VK}
v(\bK) = \frac{4\pi}{\bK^2}.
\end{equation}
Transforming to real space, through a discrete Fourier transform restricted to
one interaction cell, yields an expression that contains correction terms to
the simple $1/r$ form that will vanish in the limit of infinitesimal grid
spacing and infinite IC size. Since, in practice, we evaluate the convolution
Eq.\ (\ref{EPSREAL}) as a multiplication in reciprocal space, there is no need
to explicitly transform Eq.\ (\ref{VK}) to real space.

In practice, to deal with the divergence of the Coulomb potential
at zero wavevectors, we follow the procedure described in the
literature\cite{HYL87,BAR86}, i.~e., we employ ${\bf k}\cdot{\bf p}$
perturbation theory for calculating the head (${\bf k}={\bf G}={\bf
G}^{\prime}=0$) and wings (${\bf k}={\bf G}=0;\; {\bf G}^{\prime}\neq 0$)
of the symmetrised dielectric matrix Eq.\ (\ref{symdielm}). The matrix
elements of type $\langle \Psi_{n{\bf q}}|e^{-i \bk\cdot\br}|  \Psi_{n,{\bf
q}-{\bf k}} \rangle$ appearing in the expression for the RPA polarizability in
reciprocal-space representation (cf.\ Eq.~(18) of Ref.\ \onlinecite{HYL87})
are evaluated by expanding the wavevector dependence of the wavefunctions
and the exponential for small ${\bf k}$. This yields the lowest-order
(in ${\bf k}$) terms for head and wings of the polarizability. The head
of the polarizability goes to zero like $k^2$ thus cancelling the $1/k^2$
divergence of the Coulomb potential at ${\bf k}={\bf G}={\bf G}^{\prime}=0$,
yielding a finite value for the head of $v P^0$ whereas the lowest-order
term for the wings $P^0_{0{\bf G}^{\prime}}(0)$ is proportional to $k$,
cancelling the $1/k$ divergence of $v_{0{\bf G}^{\prime}}(0)$ and yielding
a finite $(v P^0)_{0{\bf G}^{\prime}}(0)$. Head, wings and body of the
inverse dielectric matrix are then obtained in terms of head, wings and
body of the dielectric matrix by block-wise inversion as described by
Pick, Cohen and Martin\cite{PCM70}.

The inversion of the dielectric matrix can be performed either in mixed
space or in reciprocal space. While this is strictly equivalent if the
whole FFT grid is used in either space, it is important to look at ways
to reduce the grid size without compromising the exactness of the result,
as the matrix inversion is in fact the one operation in the computational
sequence which scales worst with system size (unless sparseness is exploited,
see below) and will therefore be the bottleneck for large systems. In mixed
space we have to invert square matrices of dimension $N_{UC}$ for every $\bk$
point in the irreducible wedge of the Brillouin zone (IBZ).

First we look at the case where the IC is in no dimension smaller than the UC.
We are then dealing with the inversion of fully occupied square matrices, which
scales cubicly with the dimension of the matrix, so that the inversions
together scale like $N_{IBZ}\times N_{UC}^3$. This number is essentially the
same as $N_{IUC}\times N_{IC}\times N_{UC}$, small differences arising possibly
because of the coarseness of the grid. This shows that the scaling is
quadratic with system size for constant IC size and constant $N_{UC}/N_{IUC}$.
If the increase of the unit cell size means also a lowering of symmetry, an
additional factor corresponding to the symmetry reduction comes in.
Transforming to reciprocal space and cutting back to a set of reciprocal
lattice vectors $\bG$ within a sphere as described earlier reduces the
dimension of the matrices by a factor of 2 to 4, which speeds up the inversions
by a factor of 8 to 64, because of the cubic scaling. This is the procedure we
choose routinely in our calculations.

If the unit cell size exceeds the interaction cell size in one or more
dimensions the matrix in mixed space becomes sparse, as all those elements
for which $\br'$ is outside the IC become zero. This sparseness can be
exploited to the effect that the matrix inversion scales only as $N_{UC}^2$,
so that the overall scaling with system size remains the same. However,
in the scheme we are employing at the moment this sparseness is not yet
explicitly exploited so that we are in fact dealing with a situation where
the IC grows with the UC, leading for the matrix inversion to an $N_{UC}^3$
scaling once the unit cell outgrows the interaction cell.

\subsection{Screened Coulomb interaction}
\label{SCRCOUL}

The screened Coulomb interaction is defined in real space as
\begin{equation}
W(\br,\br';i\omega) = \int d\br''\epsilon^{-1}(\br,\br'';i\omega)v(\br''-\br')
\end{equation}
The convolution on the right hand side is again best dealt with as a
multiplication in reciprocal space 
\begin{equation}
W_{{\bf GG}^{\prime}} ({\bf k}; i\omega) = 
 \epsilon_{{\bf GG}^{\prime}} ({\bf k}; i\omega) v_{{\bf GG}^{\prime}}({\bf k}).
\end{equation}
However, in a semiconductor and in metals for frequencies other than zero, $W$ 
is truly long ranged, and therefore diverges for zero wavevectors.
In order to avoid problems (in the ${\bf G}/{\bf r}$ and 
${\bf G}^{\prime}/{\bf r}^{\prime}$  FFT) associated with the resulting 
long-range tail and (in the $i\tau/i\omega$ FFT) 
with the asymptotic frequency dependence we separate $W$ into a long-range 
part which is immediately set up in real space and a short-range part 
$W_s$ which is first computed in reciprocal space. From the long-range 
behavior of the dielectric function we find the isotropic part of the 
screened interaction in the long-range limit
\begin{equation}
\label{WLONG}
\lim_{\left|\br-\br'\right|\rightarrow\infty} W(\br,\br';i\omega)
= f(i\omega)v(\br-\br')
\end{equation}
so that we can define a short-ranged part $W_s$ of $W$
\begin{eqnarray}
W_s(\br,\br';i\omega) & = & \int d\br''\left(\epsilon^{-1}(\br,\br'';i\omega) 
                      - f(i\omega)\delta(\br-\br'')\right) \nonumber \\
                      && \hskip 25pt \times v(\br''-\br').
\end{eqnarray}

$W_s$ has well-defined Fourier coefficients for all reciprocal vectors. It
can thus be computed in reciprocal space. The long-range part is set up
in real space with $f(i\omega)$ determined from the small-wavevector behavior
of the inverse dielectric matrix in reciprocal space, i.~e., the head of the
inverse dielectric matrix $\epsilon_0^{-1}$:
\begin{equation}
\label{FIW}
f(i\omega)  = \epsilon_0^{-1}(i\omega)-1,
\end{equation}

Even though the interaction is truly long-ranged we only need to set it up
inside the IC because we aim to calculate the self-energy whose range is 
determined by that of the (short-ranged) Green's function $G_0$.  
Hence, at this point, it is sufficient to take the contribution of the 
long-range part of W into account properly {\em within} the IC. 
The Coulomb potential at $\bf r=\bf r'$ is approximated by the average 
over a sphere around $\bf r=\bf r'$ with the same volume as one real-space grid
cell.

\subsection{Self-energy}

The self-energy operator is set up in the real-space and imaginary-time domain:
\begin{equation}
\Sigma(\br,\br';i\tau) = iG^{LDA}(\br,\br';i\tau)W(\br,\br';i\tau-i\delta).
\end{equation}
Within the discrete-grid and finite-range approximation a complete description
is given if $\br$ is restricted to the IUC and $\br'$ to an IC around $\br$.
The scaling is again linear with $N_{IUC}$ for fixed $N_{IC}$.

\subsection{Expectation values of the self-energy}

To compute the expectation values of $\Sigma$ between wavefunctions at the
special $\bk$ points, we transform $\Sigma$ to mixed space and form the matrix
elements
\begin{equation}
\left<\Psi_{nk}|\Sigma(i\omega)|\Psi_{nk}\right> =
\sum_{\br,\br'}^{UC}\Psi^*_{n\bk}(\br)\Sigma_\bk(\br,\br';i\omega)
\Psi_{n\bk}(\br')
\end{equation}
This operation again scales quadratically with the number of points in the
UC for any given set of quantum numbers $n\bk$. For a general point {\bf q} in
the first Brillouin zone that is not on our discrete grid, we generate 
interpolated values through the formula
\begin{equation}
\Sigma_{\bf q} = \sum_\bR^{IC} \Sigma_\bR\exp{(-i{\bf q}\cdot\bR)}.
\end{equation}

\section{Quasiparticle energies}
\label{SECQPE}

\subsection{Analytic continuation}

In order to calculate the quasiparticle corrections to the LDA eigenvalues
we require the self-energy operator as a function of real energy, and thus
a key requirement for our imaginary time method is the ability to obtain
the expectation values of $\Sigma$ on the real energy axis accurately from
the imaginary energy behaviour. From complex analysis we know that if
two functions are equal over any arc in the complex plane then they are
equal everywhere in their common region of analyticity. We know from the
structure of $G$ and $W$ that $\Sigma(z)$ ($z$ denoting complex energy)
has poles in the second and fourth quadrant of the complex plane. If we
know the analytic form of the expectation values of the self-energy on the
imaginary energy axis we can analytically continue it from the negative
imaginary energy axis to the negative real energy axis, and from the positive
imaginary to the positive real axis, without crossing any branch cuts.

To obtain such an analytic form, we fit a multipole model
function for each pair of quantum numbers $n\bk$ to the values
$\left<\Psi_{n\bk}|\Sigma(i\omega)|\Psi_{n\bk}\right> =
\left<\Sigma(i\omega)\right>$: 
\begin{equation}
\left<\Psi_{n\bk}|\Sigma(z)|\Psi_{n\bk}\right> \simeq a^0_{n\bk} +
\sum_{i=1}^n \frac{a^i_{n\bk}}{z - b^i_{n\bk}}, \end{equation} 
where $a^i_{n\bk}$ and $b^i_{n\bk}$ are complex fit parameters and $z$ is the
complex energy. In Fig.\ \ref{fitqual} the resulting fitted form of the
correlation (energy-dependent) part of the self-energy \cite{Foot2} is 
compared with the calculated matrix elements for bands at the $\Gamma$ and $X$ 
points (the fits are plotted with the same line styles as the respective 
calculated matrix elements, they are on this scale indistinguishable from
the latter). A simple two-pole model ($n=2$) performs very successfully
for Si, with the fitted function reproducing the actual values with an rms
error of less than 0.2\%. Including several further poles in the fitted
form proves to be stable but unnecessary, although more poles are expected
to be required for systems with multiple natural energy scales. For the
analytic continuation to be valid, the fitted pole positions $b^i_{n\bk}$
should lie in the upper half plane when fitting the negative imaginary axis
and vice versa, a condition which in practice is obeyed by the optimal
parameters. The parameters $a^0_{n\bk}$ should in principle be zero, as
$\lim_{\omega\rightarrow\infty}\left<\Sigma(\omega)\right> = 0$. Allowing
a small finite value for $a^0_{n\bk}$ has proved helpful in the fits,
though, and as we are interested mainly in energies close to the Fermi
energy and not the long-range limit this is perfectly legitimate. Since
$\left<\Sigma(i\omega)\right> = \left<\Sigma(-i\omega)\right>^*$ it is
sufficient to fit only one half-axis.

Convergence of the quasiparticle energies with respect to the parameters
of the time-energy transform grid is discussed in the next section,
but it is also important to consider as a separate question the
stability of the analytic continuation procedure. In particular if we
are to achieve smooth convergence it is important that changes in the
calculated self-energy on the real axis correspond to genuine changes
in the calculated $\left<\Sigma(i\omega)\right>$, and are not simply
resulting from instabilities in the fitting procedure.  In Figures
\ref{convmax} and \ref{convdelta} we show the convergence behaviour
of a matrix element  $\left<\Sigma(i\omega)\right>$ at $\Gamma$ with
respect to number and spacing of the energy points respectively, and
the corresponding form of the analytically continued element. It can be
seen that the fitting approach is indeed stable, with the convergence of
$\left<\Sigma(\omega)\right>$ being dictated directly by changes in the
calculated  $\left<\Sigma(i\omega)\right>$.  Thus the task of converging
the quasiparticle energies is equivalent to converging the self-energy on
the imaginary axis, with very little comparative loss of accuracy in the
analytic continuation procedure itself. 

The parameter $\omega_{max}$ (describing the energy range used in the 
calculation) mainly determines the convergence of the large-energy region 
of $Im \left<\Sigma_c(i\omega)\right>$. The energy grid spacing 
$\Delta\omega$ predominantly affects the convergence of 
$\left<\Sigma_c(i\omega)\right>$ in the region close to $\omega$= 0. 
The behavior of $\left<\Sigma_c(\omega)\right>$ (on the whole real energy axis)
is more sensitive to the shape of $\left<\Sigma_c(i\omega)\right>$ in this 
region of the imaginary axis than to the large-imaginary-energy tail of the 
latter (see also next paragraph). That is why the convergence with 
respect to $\Delta\omega$ shown in Fig.\ \ref{convdelta} appears to be
better on the imaginary axis than on the real axis.

Our investigations of the convergence of the calculated matrix elements
of the correlation part of the self-energy and the resulting
quasiparticle energies with respect to $\omega_{max}$
have shown that good convergence can be achieved by keeping the energy
range for the fitting procedure fixed rather than fitting the whole range
of energies $(-\omega_{max},+\omega_{max})$ when increasing the latter.
The energy range for fitting the matrix elements $\langle \Sigma_c
(i\omega) \rangle$ was fixed to ($-$5 Har, +5 Har) in the present calculation 
(the results are not sensitive to the exact value). The reason for restricting
the fit range for $\langle \Sigma_c (i\omega)\rangle$ is that
most information is contained in the form of this function at imaginary
energies reasonably close to $i\omega = 0$.  Thus, fitting a large range,
beyond a certain point, will mean that this part of the shape will be
less accurately described due to the loss of weight in this region. This
effect prevents $\langle \Sigma_c (\omega)\rangle$ from
converging properly unless the fit range is kept fixed (at any reasonably
sensible value), as is illustrated by Fig.\ \ref{fixrange}.

\subsection{Quasiparticle corrections}

\label{QPECORR}

The quasiparticle energies are formally solutions of the equations
\begin{eqnarray}
\label{EQP}
\left(-\frac12\nabla^2 + V_{ext}(\br) + V_H(\br)\right)
\Psi_{n\bk}^{qp}(\br) && \nonumber \\
{}+\int d\br'\Sigma(\br,\br';\epsilon_{n\bk}^{qp})\Psi_{n\bk}^{qp}(\br')
& = & \epsilon_{n\bk}^{qp}\Psi_{n\bk}^{qp}(\br).
\end{eqnarray}

Because $\Sigma$ is in general a non-Hermitian operator, the eigenenergies
$\epsilon_{n\bk}^{qp}$ are complex, their real part being interpreted
as quasiparticle energy and their imaginary part as lifetime. It has been
observed\cite{HYL286,GSS88} that the wavefunctions obeying Eq.\ (\ref{EQP}) for
typical semiconductors and insulators have an almost complete overlap with the
LDA eigenfunctions which solve an equation in which the non-local self-energy
operator is replaced by a local exchange-correlation potential. This makes
it possible to find the quasiparticle energies by computing corrections to
the LDA eigenvalues in first-order perturbation theory in most cases. (For
some systems, however, the quasiparticle wavefunctions are expected to be
qualitatively different from the LDA wavefunctions, such as at a surface. The
space-time method allows the full quasiparticle wavefunctions and energies to
be calculated where required\cite{WGRN98}.) Usually the LDA Hamiltonian with
its eigenfunctions and eigenvalues is taken as the starting approximation,
with $\Sigma(\omega) - V_{xc}$ as a perturbation. Following an idea used by
Hedin\cite{HED65} for the electron gas, we shift the LDA eigenergies by a
constant $\epsilon_s$ that aligns the Fermi energies of the quasiparticles
before and after applying the $GW$ correction. This is intended to simulate
to some extent the effect of self-consistency in $G$ and has been shown in
a model system\cite{SCH97} to be instrumental in keeping charge conservation
violations negligible.

Calculation of the full energy dependence of the self-energy (via the analytic 
continuation to real energies) allows us to solve the equation
\begin{equation}
\label{QPES}
\epsilon_{n\bk}^{qp} = \epsilon_{n\bk}^{LDA} +
\left<\Psi_{n\bk}\right|\Sigma(\epsilon_{n\bk}^{qp}) -
V_{xc} - \epsilon_s\left|\Psi_{n\bk}\right>
\end{equation}
for the quasiparticle energies self-consistently. This approach was employed 
for the calculation of the quasiparticle energies in the present paper. 
Alternatively, a Taylor expansion of 
$\Sigma (\omega)$ at $\omega = \epsilon_{n\bk}^{LDA}$ can be used:
\begin{equation}
\label{QPET}
\epsilon_{n\bk}^{qp} = \epsilon_{n\bk}^{LDA} +
\frac{1}{Z_{n\bk}}\left<\Psi_{n\bk}\right|\Sigma(\epsilon_{n\bk}^{LDA}) -
V_{xc} - \epsilon_s\left|\Psi_{n\bk}\right>,
\end{equation}
\begin{equation}
Z_{n\bk} = 1 - \left.\frac{d}{d\omega}
\left<\Psi_{n\bk}\right|\Sigma(\omega)\left|\Psi_{n\bk}\right>
\right|_{\omega=\epsilon_{n\bk}^{LDA}}.
\end{equation}
Because of the analytic fit of the expectation values of the self-energy the
derivative with respect to the energy argument is readily found in closed form.
For $\epsilon_s$ a closed form but rather involved expression can also be
given. Comparison of the self-consistent solution of Eq.\ (\ref{QPES}) and the 
result of the Taylor expansion  Eq.\ (\ref{QPET}) shows that the latter is 
usually a very good approximation, yielding quasiparticle energies differing 
by only up to 5 meV from the solution of Eq.\ (\ref{QPES}) for bulk Si.

\section{Results for bulk silicon and silicon supercells}
\label{RESULTS}

In order to test the performance and the scaling properties of the
space-time $GW$ method we have performed calculations of self-energy and
quasiparticle energies for bulk silicon and silicon with artificially
increased unit cell sizes. The latter were slab-like supercells containing
four and eight atoms, respectively, with tetragonal symmetry, i.~e., the
unit cells were artificially enlarged in one direction. Supercells
of this geometry can be used to model semiconductor superlattices and
surfaces.

\subsection{Convergence parameters}

Before discussing the performance and scaling issues we briefly summarize
the results of convergence tests for bulk Si. In Table \ref{siconv}
the parameters needed to converge quasiparticle energy differences to
an accuracy of 50 meV and 20 meV, respectively, with the present method
are gathered. This accuracy is meant to be with respect to each parameter
individually. We estimate the overall accuracy of our results to be better
than 0.1 eV. The plane-wave (PW) cutoff in Table \ref{siconv} is the cutoff
of $\frac{1}{2}({\bf k}+{\bf G})^2$ in the LDA calculation providing the
wavefunctions $\Psi_{n{\bf k}}$. All PW components of the LDA wave functions
are used in the $GW$ calculation. The real-space grids in the unit cell
(see Section \ref{DISCRETIS}) resulting from PW cutoffs of 13.5 Ry and 16
Ry comprise $9\times9\times9$ and $12\times12\times12$ points, respectively.
The band cutoff determines the number of bands included in the band summation
in Eq.~(\ref{GLDAST}) for the calculation of the Green's function.  Band
cutoffs of 8 and 10 Ry correspond to taking 109 and 145 bands, respectively,
into account. For sampling the BZ we employ a regular non-offset grid of
${\bf k}$ points. The LDA wavefunctions are calculated at the ${\bf k}$ points
in the irreducible wedge of the BZ. As described in Section \ref{DISCRETIS}
the dimensions of the ${\bf k}$ grid in the BZ are equal to the dimensions
(in multiples of primitive lattice vectors) $N_i^{\bf R}$ of the IC, which
must be large enough to comprise the range of non-locality of the Green's
function and the self-energy in Si. The smaller ${\bf k}$ grid given in Table
\ref{siconv} can roughly accomodate a non-locality range of 14.5 a.u.
and the larger of 18 a.u. $(-\tau_{max},+\tau_{max})$ and $\Delta \tau$
stand for the range and spacing of the equidistant grid for sampling the
functions $F({\bf r},{\bf r}',i\tau)$ on the imaginary time axis. The
parameters $\omega_{max}$ and $\Delta\omega$ of the corresponding grid
in the imaginary energy domain are related to the time-grid parameters by
$\Delta\omega = 2\pi/(2N_{\tau}-1) \Delta \tau$ with $\tau_{max} = N_{\tau}
\Delta \tau$.

\subsection{Quasiparticle energies for bulk silicon}

The quasiparticle energies for bulk Si obtained from a well
converged calculation with the present method are given in Table
\ref{siresult}\cite{Foot3}. The LDA calculation was done with the PW
pseudopotential method employing a pseudopotential constructed according
to the prescription of Hamann\cite{HAM89} and using the Ceperley-Alder
exchange-correlation potential\cite{CEA80} in the parametrization of
Perdew and Zunger\cite{PEZ81}. As can be seen from Table \ref{siresult}
the calculated $GW$ quasiparticle energies agree well with quasiparticle
excitation energies derived from photoemission, inverse photoemission
and optical experiments\cite{LAB82,ORH93,SED68,WMH85,HHE81,HUN76,SLH85},
except for the position of the bottom of the valence band which appears to
be too high in our calculation. The agreement of the quasiparticle energy
differences obtained with the present method ($GW$ 1 in Table \ref{siresult})
with those of a recent $GW$ calculation by Fleszar and Hanke\cite{FLH97}
is very good. Like us, these authors did not make use of a plasmon pole
model to describe the energy dependence of the dynamically screened Coulomb
interaction but directly computed its full energy dependence.

We investigated how the quasiparticle energies change when the Green's
function is updated by replacing the LDA eigenvalues by the quasiparticle
energies and the self-energy recalculated with the updated Green's function
as input, as was done by Hybertsen and Louie \cite{HYL286}. The results
($GW$ 2) are also included in Table \ref{siresult}. Updating the Green's
function increases the gap by about 0.1 eV. Now the conduction band energies
agree within better than 0.1 eV with those given in Ref.\ \onlinecite{HYL286}
whereas the valence band energies remain higher by between 0.1 eV and 0.5 eV
(relative to the top of the valence band), the difference increasing with
the distance of the respective state from the Fermi level.

Indeed, the results of several earlier $GW$ calcula\-tions\cite{HYL286,RKP93}
are somewhat different from those given in Table \ref{siresult} (and Ref.\
\onlinecite{FLH97}). This can be attributed to the following reasons:
\begin{enumerate}
\item[(i)] a possible lack of convergence with respect
to the number of conduction bands taken into account in the earlier
calculations, causing the topmost occupied state (which is particularly
sensitive to this parameter) to be about 0.2 eV too high, as discussed in
Ref.\ \onlinecite{FLH97}; and
\item[(ii)] the use of plasmon pole models
for the energy dependence of the dynamically screened Coulomb interaction
in the earlier calculations.
\end{enumerate}

\subsection{Silicon supercells}

In Table \ref{cellresult} we compare the quasiparticle energies for
slab-like Si$_4$ and Si$_8$ supercells to the bulk Si results. Shown are
the quasiparticle energies for the high-symmetry ${\bf k}$ points of Si
(which are not necessarily symmetry points for the supercell geometry)
and for those ${\bf k}$ points which are mapped onto these symmetry points
when the unit cell size is increased. The calculations were done with a PW
cutoff of 13.5 Ry (leading to a real-space grid spacing of $\Delta r$ = 0.8
a.u.), a band cutoff of 10 Ry, $\Delta \tau$ = 0.3 a.u. and $\tau_{max}$ =
20 a.u. The ${\bf k}$ grids have been chosen to make the IC of equal size
in all three dimensions for all the calculations, for the reason discussed
in Section \ref{DISCRETIS}. The grid sizes resulting from these parameters
are summarized in Table \ref{gridpar}. $N_{IUC}$, $N_{UC}$, and $N_{IC}$
are the numbers of real-space grid points in the irreducible part of
the unit cell, the unit cell and the interaction cell, respectively (see
Section \ref{ASCALING} above). $N_{k}$ stands for the number of ${\bf k}$
points in the BZ and $N_{t/\omega}$ is the number of positive time/energy
points. The supercell results in Table \ref{cellresult} agree with
the bulk Si results, as expected. The small differences for the highest 
conduction band given in Table \ref{cellresult} are ascribed to the lower 
symmetry of the supercells leading to different ${\bf G}$ vector sets.

\subsection{Scaling}

As outlined in Section \ref{ASCALING} the computational effort in the present
method should scale quadratically with the number of atoms in the unit
cell.  Fig.\ \ref{scaling} shows the CPU times for a full $GW$ calculation
for Si$_n$ ($n$ = 2, 4, 8) on a Digital Alpha 500/500 workstation. The
parameters of these calculations are those given in Table \ref{gridpar}
and the discussion thereof.  Si$_2$ (bulk Si) has a higher symmetry than
the tetragonal supercells Si$_4$ and Si$_8$. As the symmetry is exploited
in most parts of the calculation this saves approximately a factor of two
in CPU time (cf.~$N_{IUC}$ values in Table \ref{gridpar}). Taking this into
account the overall scaling is indeed quadratic, as expected. The most time
consuming parts of the calculation are:
\begin{enumerate}
\item[(i)] 
setting up the Green's function according to Eq.\ (\ref{GLDAST})
\item[(ii)] 
calculating the dynamically screened Coulomb interaction (including the
inversion of the dielectric matrix \cite{Foot4}, the transformation to and from
reciprocal space, and reading of $P^0$ from and writing of $W$ to disk)
\item[(iii)] 
computation of the matrix elements of the energy dependent self-energy
for a given number of ${\bf k}$ points and bands.
\end{enumerate}
Optionally, the Green's function can be recomputed when the self-energy is
calculated, rather than storing it on disk when it is first set up
and reading it in for the calculation of the self-energy. 
This procedure reduces the required disk space by
a factor of 2/3. It has been used for the Si supercell calculations
described here. As can be seen from the breakdown of the total CPU time
shown in Fig.\ \ref{scaling} all these parts of the calculation scale roughly
like $N^2$. More precisely, the scaling is like $N_{IUC} N_{bands}$ for
(i), $N_{IUC} N_{UC}$ for (ii) and $N_{UC}^2$ for (iii) as was discussed
in Section \ref{ASCALING}. 
%When looking at the total computational times
%it should be borne in mind that the full self energy 
%$\Sigma({\bf r},{\bf r}',i\omega)$ is calculated rather than just a number 
%of matrix elements 
%$\left<\Psi_{n\bk}\right|\Sigma(\omega)\left|\Psi_{n\bk}\right>$.

Comparing the performance of our method with conventional techniques we
note that although a number of quasiparticle calculations within the GW
approximation have been reported for systems with up to 60 atoms,
e.~g.~for surfaces \cite{HyL87s,Northrup93s,Rohlfing96s}, defects 
\cite{Surh95}, and fullerenes \cite{ShiL93f}, several simplifying approximations 
have been employed in these calculations in order to reduce the computational 
effort. These authors focus on the calculation of quasiparticle energies and 
therefore only a number of self-energy matrix elements were computed. 
This is in contrast to the present method which provides the full 
self-energy (thus giving acccess to quantities other than the quasiparticle
energies, see conclusions section below). In all the calculations mentioned 
above plasmon-pole models have been employed to approximate the energy
dependence of the dynamically screened Coulomb interaction. In some cases 
\cite{Northrup93s,ShiL93f} a model static dielectric matrix has been used
as well. To aid comparison we note that although the plane-wave 
cutoff of the underlying LDA pseudopotential calculation can be (and has been 
in many GW calculations reported in the literature) considerably reduced in the 
GW calculation because the quasiparticle corrections usually converge much 
faster with this parameter than the LDA eigenvalues, this has
not been done in the test calculations reported in the present paper.
The energy dependence of the screened interaction is fully taken into
account in the method proposed in this paper. Hence the dielectric matrix 
is computed and inverted for about 60 to 100 imaginary energies which is
computationally much more demanding than using a plasmon-pole model.
We estimate that the crossover to the advantegeous scaling behavior of CPU time
in our method occurs (a) already for bulk Si if we compare with a calculation of
the full self-energy with a conventional reciprocal-space approach and (b) for
%a system with about 20 atoms if we compare with a reciprocal-space method
systems in the range studied in this paper if we compare with a 
reciprocal-space method
where a plasmon-pole model is used and which is restricted to computing a
moderate number of self-energy matrix elements only. Finally, we mention
that work, particularly regarding the treatment of the time/energy 
dependence of the key quantities, is in progress to reduce the prefactor of 
the scaling of our method.

\section{Summary}

We have presented a detailed account of a method for calculating the electron
self-energy and related quantities from ab-initio many-body perturbation
theory within the $GW$ approximation. The method is based on representing the
basic quantities (Green's function, dielectric response function, dynamically 
screened Coulomb interaction and self-energy) on a real-space grid and on
the imaginary time axis. In those intermediate steps of the calculation where
it is computationally more efficient to work in reciprocal space and
imaginary energy we change to the latter representation by means of fast
Fourier transforms. Working on the imaginary time/energy axis considerably
facilitates the numerical treatment. The matrix elements of the self-energy
on the real-energy axis are obtained by an analytic continuation procedure
which was shown to be accurate and stable. We have demonstrated the accuracy of
the method by calculating quasiparticle excitation energies at high-symmetry
points for the prototype semiconductor Si. The computational effort of the
method scales quadratically with unit cell size. This was shown
explicitly by calculations for Si supercells. The method allows the
extension of ab-initio work beyond the calculation of quasiparticle energies
and its application to materials with larger basis sets or larger unit cells
than were previously feasible. Calculating the full self-energy gives access
to quantities like the charge density\cite{RIG98}, spectral function
\cite{RGN95}, momentum distribution and total energy at the $GW$ level. It 
also opens the possibility of calculating the self-energy self-consistently and 
to provide the $GW$ Green's function (after solving the Dyson equation) -- 
essential prerequisites for going beyond the $GW$ approximation, e.~g., by 
iterating Hedin's equations. This remains to be investigated in future.

\section{Acknowledgments}
We acknowledge helpful discussions with R.~J. Needs and A. Rubio.  This work was 
supported by the Engineering and Physical Sciences Research Council.  
M.~M. Rieger acknowledges financial support by the European Union 
under the TMR scheme.

%\bibliography{papers,foot}

\begin{references}
\bibitem{Foot1}
Note that the signs of the exponents and of $\tau$ were given incorrectly in
  Eq.~(3) of Ref.\ \onlinecite{RGN95}.
\bibitem{Foot2}
We define this quantity as the difference between the GW self-energy and the
  energy-independent bare-exchange part of the self-energy.
\end{references}
%\bibliographystyle{prsty}

\begin{figure}
\caption{The grids and cells used in the calculation in (a) real and (b) 
reciprocal space. These are shown here schematically in two dimensions for the 
case of a $2 \times 2$ grid of {\bf k}-points (corresponding to an interaction 
cell of $2 \times 2$ unit cells), and a $3 \times 3$ real-space grid in the unit 
cell (corresponding to a $3 \times 3$ grid of reciprocal lattice vectors in 
reciprocal space). The bare Coulomb interaction 
$1/ \left| {\bf r}-{\bf r}^{\prime }\right| $ is non-periodic on the interaction 
cell, and its amplitude at a real-space-grid point is taken to be that at the 
corresponding point in the Wigner-Seitz cell around $\br$ of the IC lattice. 
All other quantities, in both real and reciprocal space, are periodic. 
In these cases the choice of the primitive (shown) or Wigner-Seitz cell is a 
matter of computational convenience.}
\label{grids}
\end{figure}

%above caption goes with grid.ps

\begin{figure}
\caption{Fitting of the matrix elements of the correlation self-energy 
$\langle \Sigma_c(i\omega) \rangle$ of silicon as a function 
of imaginary energy. The real (top panel) and imaginary (lower panel) parts are 
both reproduced extremely accurately by the fits (which on this scale are 
indistinguishable from the calculated values). The bands shown are band 1 at 
the $\Gamma$ point (solid line for fitted and calculated function), 
bands 2-4 at $\Gamma$ (dotted line), bands 3-4 at $X$ (dashed) and bands 5-6 
at $X$ (dot-dashed).}
\label{fitqual}
\end{figure}

%above caption goes with sigiw.conv.ps

\begin{figure}
\caption{Convergence of a matrix element (degenerate bands 2-4 at $\Gamma$) of 
the correlation self-energy $\langle \Sigma_c(i\omega) \rangle$ 
of silicon with respect to $\omega_{max}$ with a fixed energy grid spacing of 
$\Delta\omega$= 0.16 Har. The top panel shows the calculated self-energy on 
the imaginary axis with the analytically continued dependence on real energy 
shown in the lower panel. The lines correspond to  $\omega_{max}$= 5 Har 
(dotted), 10 Har (dashed) and 20 Har (solid). 
Changes in $\langle \Sigma_c(\omega)\rangle$ are directly related to changes in 
the calculated $\langle \Sigma_c(i\omega)\rangle$, rather than any
instabilities in the analytic continuation technique.
}
\label{convmax}
\end{figure}

%above caption goes with sig.wmax.ps

\begin{figure}
\caption{Convergence of $\langle \Sigma_c(i\omega) \rangle$ with respect to 
energy grid spacing $\Delta \omega$ with fixed $\omega_{max}$= 10 Har.  
The lines correspond to  $\Delta \omega$= 0.24 Har (dotted), 0.16 Har 
(dashed) and 0.08 Har (solid). As in the previous Figure the analytically 
continued matrix element converges well, indicating the stability of the
fitting procedure. The matrix elements on the real-energy axis (which are
obtained in form of fits to a model function, see text) are plotted with a 
fixed grid spacing of 0.04 Har in order to facilitate comparison between the 
three curves.
}
\label{convdelta}
\end{figure}

%above caption goes with sig.dw.ps 

\begin{figure}
\caption{Real part of a matrix element (degenerate bands 2-4 at $\Gamma$)
of the correlation self-energy $\langle \Sigma_c (\omega) \rangle$ on the 
real energy axis, computed with $\omega_{max}$ = 5 Har (dot-dashed line), 
10 Har (dashed line), and 20 Har (long-dashed line) with fitting 
$\langle \Sigma_c (i\omega) \rangle$ on the imaginary energy axis using 
the energy range $(-\omega_{max},+\omega_{max})$
and with $\omega_{max}$ = 10 Har (dotted line) and 20 Har (solid line)
using a fixed energy range of ($-$5 Har, +5 Har) for the fitting. 
}
\label{fixrange}
\end{figure}

%above caption goes with sigwfix.ps 

\begin{figure}
\caption{Scaling of the CPU time on a Digital Alpha 500/500 workstation
with respect to unit cell size for Si$_n$ ($n$ = 2,4,8). Besides the total
CPU time ($\circ$) the times for three major parts of the computation
are given: calculation of (i) the Green's function ($\Box$), (ii) the
dynamically screened Coulomb interaction ($\Diamond$), and (iii) the matrix
elements of the self-energy including a recomputation of the Green's function
($\triangle$), see text. Note the quadratic spacing of the abscissa.
}
\label{scaling}
\end{figure}

%above caption goes with scalingfq.ps 

\begin{table}
\caption{Convergence of quasiparticle energies for bulk Si with
respect to cutoff and grid parameters in the present method. Shown are the
parameters needed for a convergence of 50 meV and 20 meV, respectively.}
\begin{tabular}{lcc}
 parameter &  50 meV &  20 meV \\
\tableline
 plane-wave cutoff (in Ry) & 13.5 &  16  \\
 band cutoff (in Ry) & 8 &  10  \\
 size of {\bf k} grid & $4\times4\times4$ &  $5\times5\times5$  \\
 spacing $\Delta t$ of time grid (in a.u.) & 0.3 &  0.15  \\
 range $t_{max}$ of time grid (in a.u.) & 13. &  20.  \\
\end{tabular}
\label{siconv}
\end{table}

\begin{table}
\caption{Calculated quasiparticle energies at points of high symmetry for
Si (in eV). The valence band maximum has been set to zero. The
calculation was done with a plane-wave cutoff of 16 Ry, a $5\times5\times5$
k-point grid, a band cutoff of 10 Ry and a time grid with $\Delta \tau$ = 0.3
a.u. and $\tau_{max}$ = 20 a.u. For comparison the eigenvalues obtained in the
LDA calculation providing the input for the $GW$ calculation have been
included. $GW$ 1 refers to a $GW$ calculation using the LDA Green function
whereas in $GW$ 2 the Green function has been updated by replacing the LDA
eigenvalues by the quasiparticle energies obtained in $GW$ 1.  In the last
column some experimental data are given (see text).}
\begin{tabular}{lrrrc}
 band &  LDA &  $GW$ 1 & $GW$ 2 & Experiment \\
\tableline
 $\Gamma_{1v}$  &  -11.89 &  -11.57 & -11.58 & $-12.5\pm 0.6$\tablenotemark[1]\\
 $\Gamma_{25v}'$ &    0.00 &    0.00 &   0.00 &   0.00 \\
 $\Gamma_{15c}'$ &    2.58 &    3.24 &   3.32 &  3.40\tablenotemark[1], 
                                                 3.05\tablenotemark[2]\\
 $\Gamma_{2c}'$  &    3.28 &    3.94 &   4.02 & 4.23\tablenotemark[1], 
                                                4.1\tablenotemark[2] \\[2mm]
 $X_{1v}$       &   -7.78 &   -7.67 & -7.68 & \\
 $X_{4v}$       &   -2.82 &   -2.80 & -2.81 & -2.9\tablenotemark[3], 
                                              $-3.3\pm 0.2$\tablenotemark[4]\\
 $X_{1c}$       &    0.61 &    1.34 & 1.42 & 1.25\tablenotemark[2] \\
 $X_{4c}$       &   10.11 &   10.54 & 10.63 & \\[2mm]
 $L_{2v}'$      &   -9.57 &   -9.39 & -9.40 & $-9.3\pm 0.4$\tablenotemark[1]\\
 $L_{1v}$       &   -6.96 &   -6.86 & -6.88 & $-6.7\pm 0.2$\tablenotemark[1] \\
 $L_{3v}'$      &   -1.17 &   -1.17 & -1.17 & $-1.2\pm 0.2$\tablenotemark[1],
                                               -1.5\tablenotemark[5] \\
 $L_{1c}$       &    1.46 &    2.14 & 2.22 & 2.1\tablenotemark[6], 
                                             $2.4\pm 0.15$\tablenotemark[7] \\
 $L_{3c}$       &    3.33 &    4.05 & 4.14 & $4.15\pm 0.1$\tablenotemark[7] \\
 $L_{2c}$       &    7.71 &    8.29 & 8.39 & \\[2mm]
 $E_{gap}$      &    0.49 &    1.20 & 1.28 & 1.17\tablenotemark[1] \\
\end{tabular}
\label{siresult}
\tablenotetext[1]{Ref.\ \onlinecite{LAB82}.}
\tablenotetext[2]{Ref.\ \onlinecite{ORH93}.}
\tablenotetext[3]{Ref.\ \onlinecite{SED68}.}
\tablenotetext[4]{Ref.\ \onlinecite{WMH85}.}
\tablenotetext[5]{Ref.\ \onlinecite{HHE81}.}
\tablenotetext[6]{Ref.\ \onlinecite{HUN76}.}
\tablenotetext[7]{Ref.\ \onlinecite{SLH85}.}
\end{table}

\begin{table}
\caption{Grid parameters for the Si$_n$ ($n$ = 2, 4, 8) calculations (see
text).}
\begin{tabular}{lccc}
 parameter               &  Si$_2$     &  Si$_4$   & Si$_8$ \\
                         &  (bulk Si)  &           &  \\
\tableline
 $N_{IUC}$               &  55         & 230       & 455 \\
 $N_{UC}$  &  $9\times9\times9$   & $9\times9\times18$  & $9\times9\times36$ \\
 $N_{k}$   &  $4\times4\times4$   & $4\times4\times2$   & $4\times4\times1$ \\
 $N_{IC} = N_{UC} N_{k}$ & $36\times36\times36$ & $36\times36\times36$ & 
$36\times36\times36$\\
 $N_{t/\omega}$          &  63         & 63        & 63 \\
\end{tabular}
\label{gridpar}
\end{table}

\begin{table}
\caption{Comparison of quasiparticle energies (in eV) at high symmetry
points for bulk Si with the corresponding values obtained from
calculations for Si$_4$ and Si$_8$ supercells (see text). All energies
refer to the respective valence band top. Shown are the band energies for
all k-points mapping onto $\Gamma$, $X$, and $L$, respectively,
when the unit cell size is increased, where those k-points appear in the
calculation. }
\begin{tabular}{lrrrrrrrrrr}
         &  \multicolumn{4}{c}{bulk Si} & \multicolumn{3}{c}{Si$_4$} &
            \multicolumn{3}{c}{Si$_8$} \\
\tableline
$\Gamma$ & -11.55& -7.63& -10.53 && -11.53 & -7.61 & & -11.55 & -7.63 & -10.53\\
         &   0.00& -2.78&  -3.41 &&   0.00 & -2.78 & &   0.00 & -2.78 &  -3.41\\
         &   3.23&  1.35&  -1.81 &&   3.22 &  1.37 & &   3.22 &  1.37 &  -1.81\\
         &   3.97& 10.52&   1.77 &&   3.96 & 10.45 & &   3.96 & 10.43 &  1.78\\
         &       &      &   3.82 &&        &       & &        &       &  3.82\\
         &       &      &   6.35 &&        &       & &        &       &  6.30\\
[2mm]
$L$      &  -9.35 & -9.62 &        & &  -9.34 &      & &  -9.36 & -9.63 & \\
         &  -6.83 & -5.35 &        & &  -6.81 &      & &  -6.83 & -5.35 & \\
         &  -1.16 & -3.62 &        & &  -1.16 &      & &  -1.16 & -3.62 & \\
         &   2.17 & -1.27 &        & &   2.18 &      & &   2.18 & -1.27 & \\
         &   4.04 &  3.13 &        & &   4.03 &      & &   4.02 &  3.12 & \\
         &   8.28 &  5.46 &        & &   8.26 &      & &   8.25 &  5.42 & \\
         &        &  5.66 &        & &        &      & &        &  5.63 & \\
         &        &  6.75 &        & &        &      & &        &  6.70 &\\[2mm]
$X$      &  -7.63 & -7.46 &        &&  -7.61 &       &&  -7.64 & -7.47 & \\
         &  -2.78 & -3.73 &        &&  -2.78 &       &&  -2.79 &  3.73 & \\
         &   1.35 &  4.84 &        &&   1.37 &       &&   1.36 &  4.82 & \\
         &  10.52 &  5.67 &        &&  10.44 &       &&  10.42 &  5.63 & \\
\end{tabular}
\label{cellresult}
\end{table}

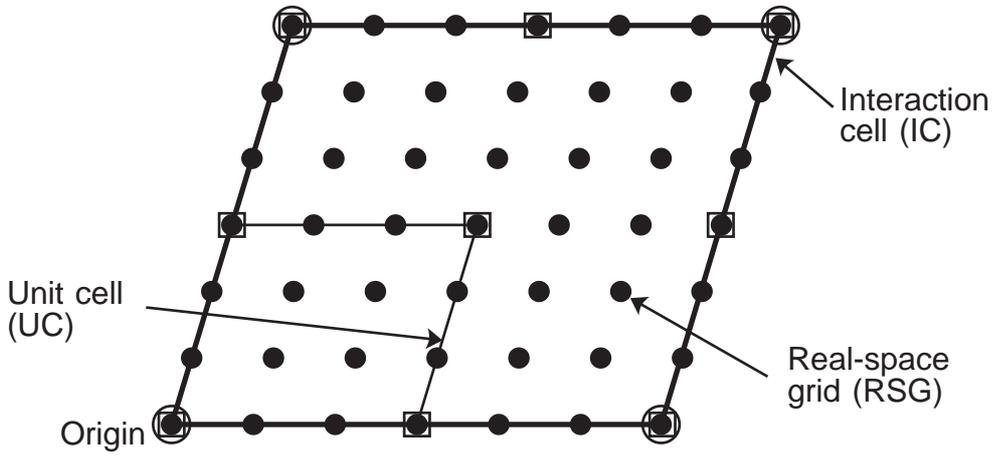
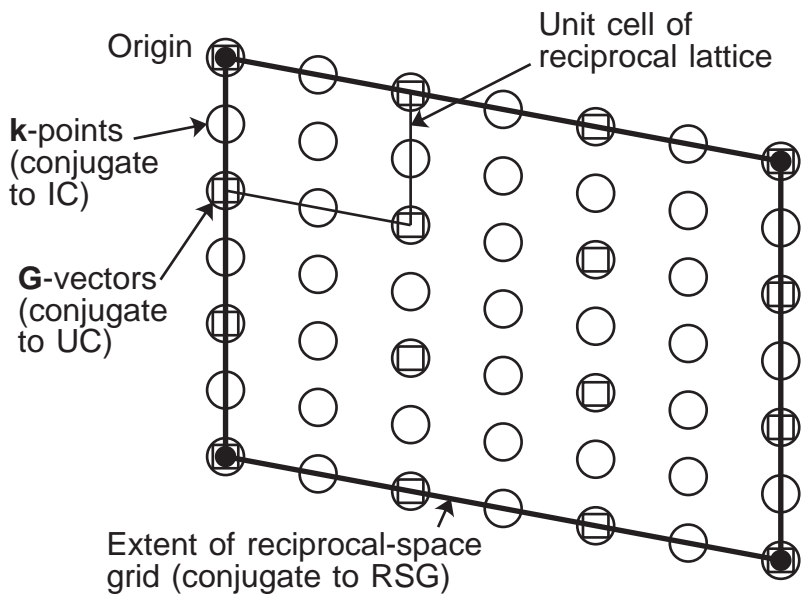
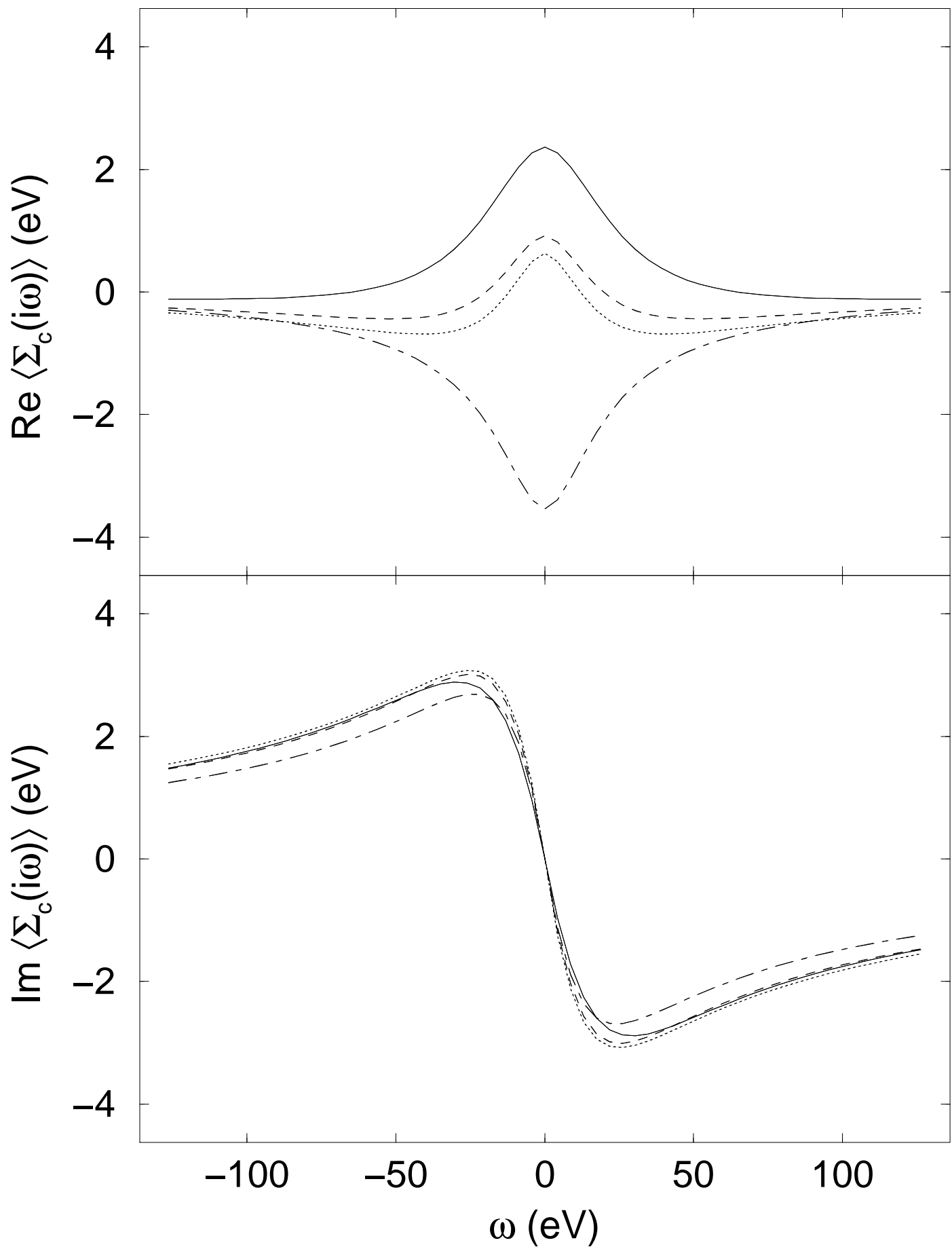
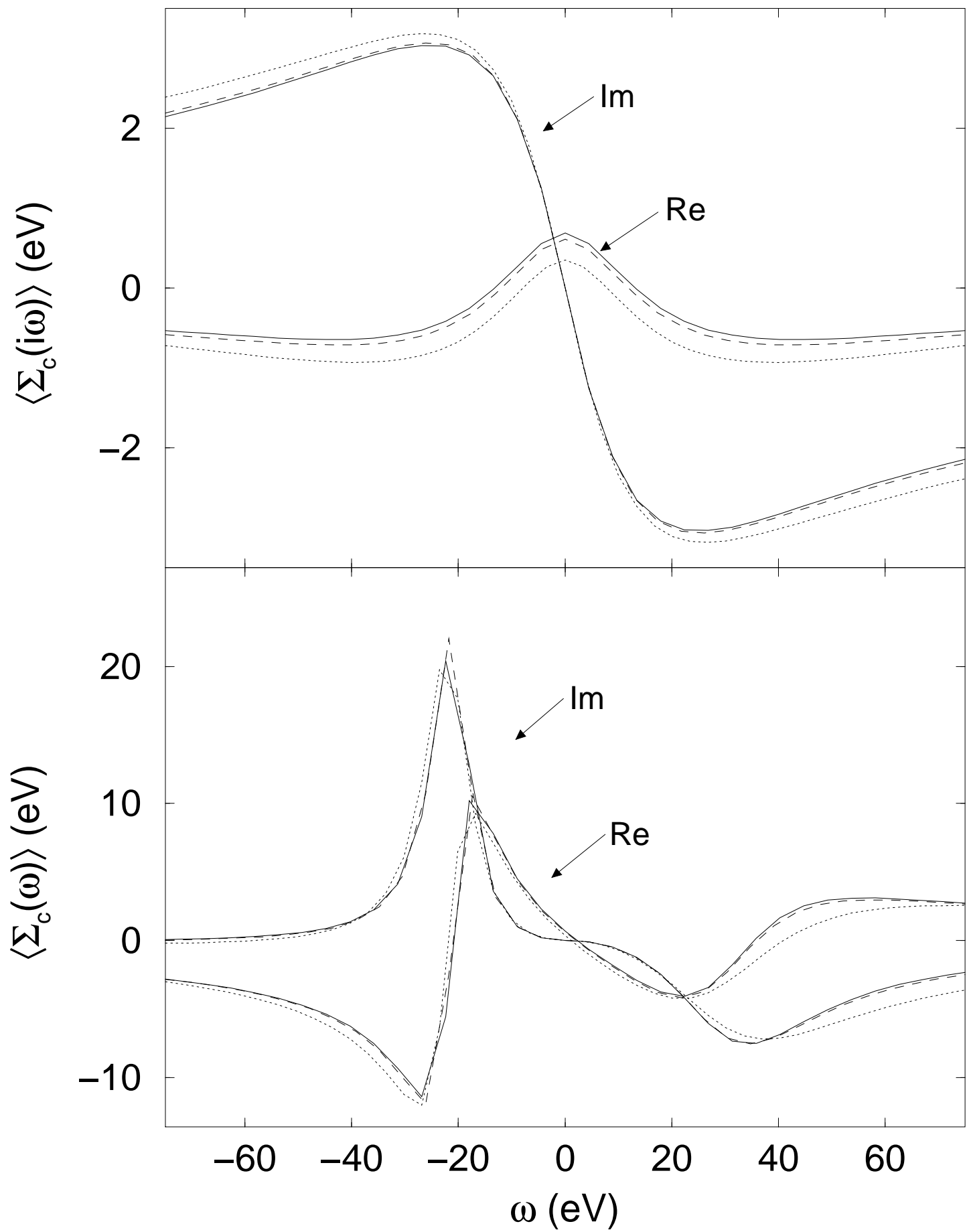
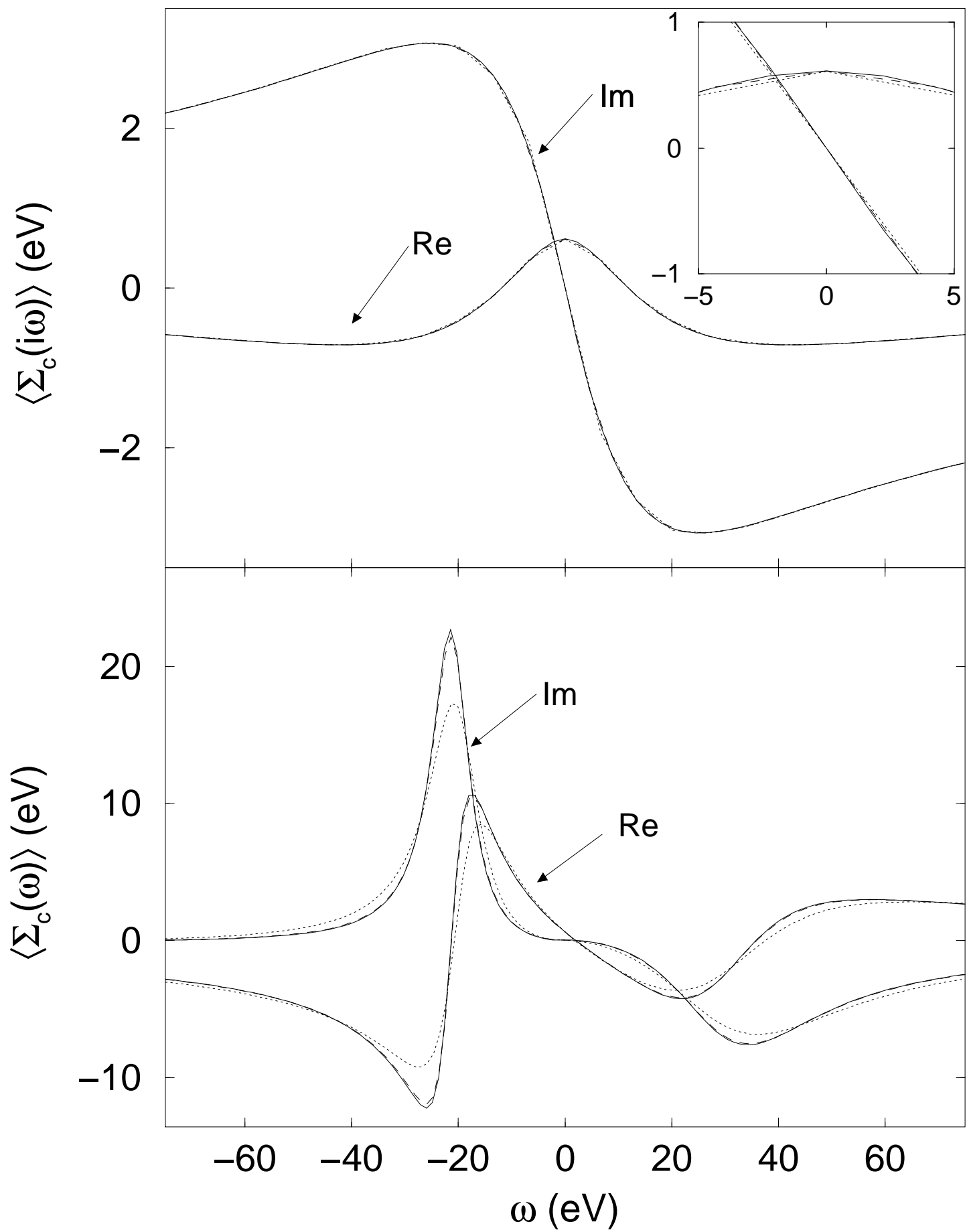
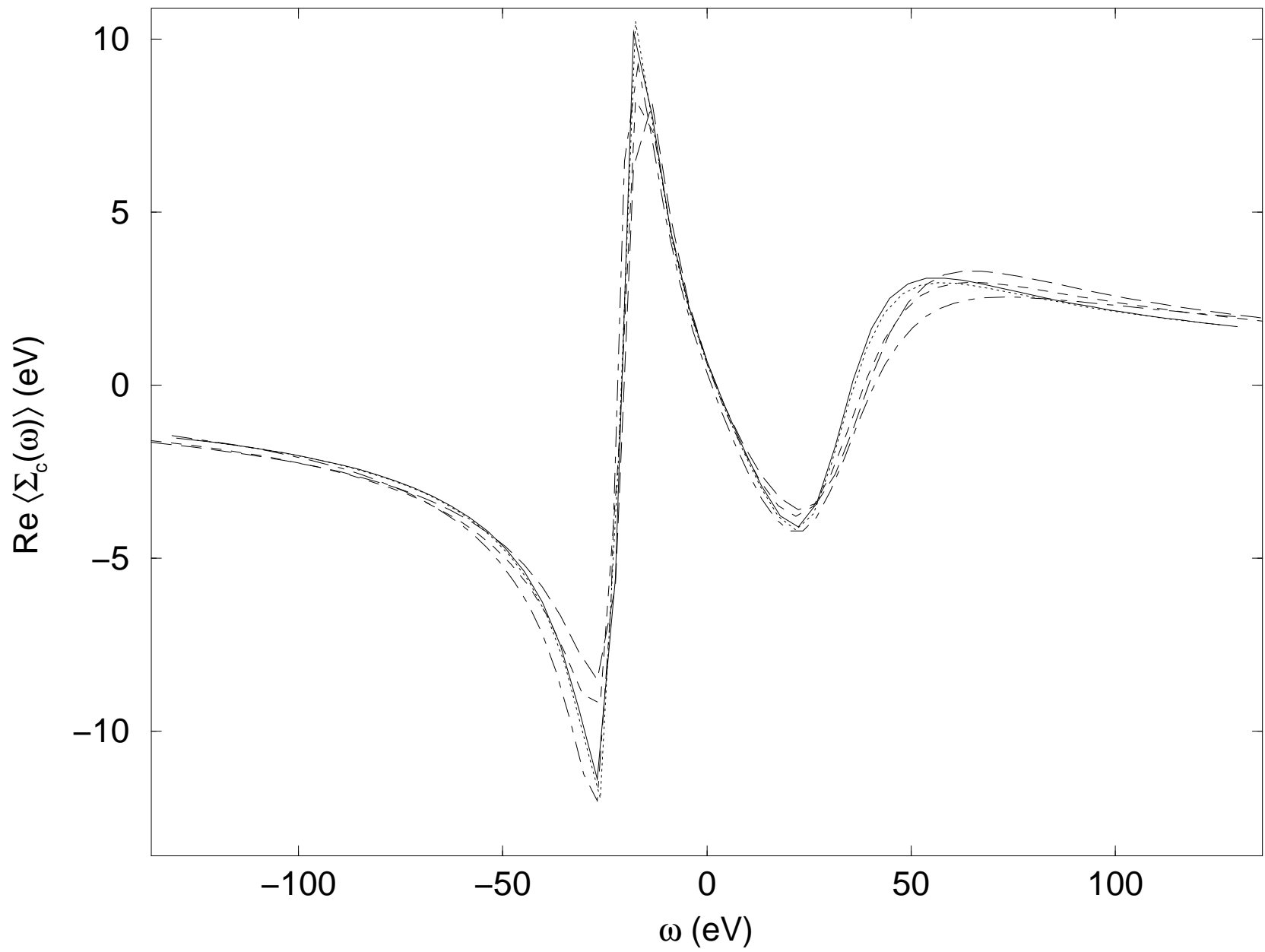
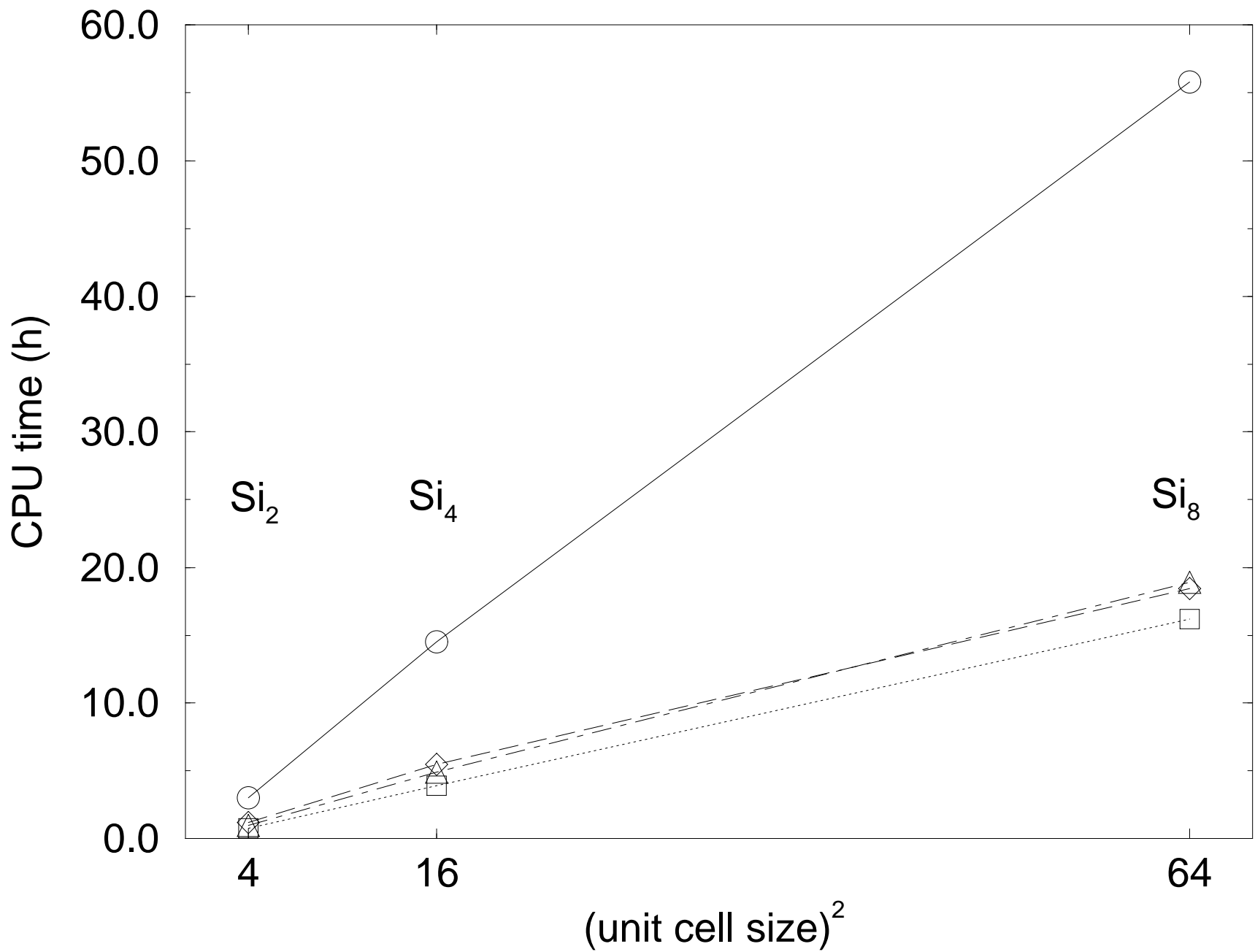
\end{document}